\documentclass[preprint,aps]{revtex4}
\usepackage{epsf}
\usepackage{graphicx} 
\usepackage{color} 

\begin{document}

\title
{
High-Order Coupled Cluster Method (CCM) Formalism 3:  Finite-Size CCM 
}

\author
{
D. J. J. Farnell
}
\affiliation
{Health Methodology Research Group, School of Community-Based Medicine, 
Jean McFarlane Building, University Place, University of Manchester,
Manchester M13 9PL, United Kingdom}

\date{\today}

\begin{abstract}
Recent developments of high-order CCM have been to extend existing formalism 
and codes to $s \ge \frac 12$ for both the ground and excited states, and independently
to ``generalised'' expectation values for a wide range of  one- and two-body spin operators.
An advantage of the CCM is that the Goldstone linked-cluster theorem
is obeyed at all levels of approximation and so it provides results in the infinite 
lattice limit $N \rightarrow \infty$ from the outset. However, recent results have 
also shown that the CCM can provide exact (symmetry-breaking)
results for  the spin-half linear-chain $J_1$--$J_2$ at the  Majumdar-Ghosh
point $J_2/J_1=0.5$ by identifying special solutions of the CCM equations for the usual
N\'eel model state. Interestingly, the CCM provides exact (non-symmetry-breaking) results 
for systems in which small magnetic clusters become de-coupled from each other when the bonds 
connecting them tend to zero. These exact results involve the identification of ``special solutions''
of the CCM equations for the N\'eel state. An example of this is given by a spin-half system with 
nearest-neighbour bonds for an underlying lattice corresponding to the magnetic material 
CaV$_4$O$_9$ (CAVO) in which one of the two different types of bonds on the lattice tend 
to zero. Larger finite-sized systems may be considered 
by appropriate choice of the unit cell and the bonds on it. We show 
here that exact diagonalisation results for ground-state energy and excitation energy
gap for the spin-half and spin-one linear Heisenberg model on chains of length up 
to $N=12$ sites for $s=1/2$ and $N=6$ sites for $s=1$ 
with periodic boundary conditions are reproduced exactly using high-order CCM
via this ``brute-force'' approach; i.e., one in which none of the translational or 
point-group symmetries of the finite lattice are used. 
\end{abstract}

\maketitle

\section{Introduction}

The coupled cluster method (CCM) \cite{refc1,refc2,refc3,refc4,refc5,refc6,refc7,refc8,refc9}  is a well-known method of quantum many-body theory (QMBT). The CCM has been applied with much success  in order to study quantum magnetic systems at zero temperature (see Refs. \cite{ccm1,ccm2,ccm999,ccm3,ccm4,ccm5,ccm6,ccm7,ccm8,ccm9,ccm10,ccm11,ccm12,ccm13,ccm13a,ccm14,ccm15,ccm16,ccm17,ccm18,ccm19,ccm19a,ccm20,ccm21,ccm22,ccm23,ccm24,ccm24a,ccm26,ccm27,ccm27a,ccm28,ccm29,ccm30,ccm31,ccm32,ccm33,ccm34,ccm35,ccm36,ccm37,ccm38,ccm39}). In particular, the use of computer-algebraic implementations \cite{ccm12,ccm15,ccm20,ccm39} of the CCM has been found to be very effective with respect to these spin-lattice problems. 
Recent developments of high-order CCM formalism and codes have been to treat systems 
with spin quantum number of $s \ge \frac 12$ for both the ground and excited states \cite{ccm39}.
Furthermore, the ground-state formalism and codes may also be used directly 
to find ``generalised'' expectation values \cite{ccm40}. These expectation values are defined 
for a wide range of one- or two-body spin operator that prior to the CCM calculation.

Here we show how the consideration of previous results for exact results for the 
(symmetry-breaking) 1D $J_1$--$J_2$ model at $J_2/J_1=0.5$ \cite{ccm37,ccm41}
and a (non-symmetry-breaking) nearest-neighbour CAVO model \cite{ccm37} 
in the limits that various nearest-neighbour bond either go to 
zero or infinity leads on naturally to the treatment of finite lattices via the CCCM code
\cite{thecode}. This is achieved by the simple expedient of choosing the finite-lattice
to be the fundamental unit cell and so this is a ``brute-force'' solution of the finite-lattice 
problem via high-order CCM. 

\section{Method}

The details of the practical application of high-order coupled cluster method (CCM) formalism
to lattice quantum spin systems  are given in Refs. \cite{ccm12,ccm15,ccm20,ccm26,ccm39} and
also in the appendices to this article. However, we point out now that the 
ket and bra ground-state energy eigenvectors, $|\Psi\rangle$ and  $\langle\tilde{\Psi}|$, 
of a general many-body system described by a Hamiltonian $H$, are given by
\begin{equation} 
H |\Psi\rangle = E_g |\Psi\rangle
\;; 
\;\;\;  
\langle\tilde{\Psi}| H = E_g \langle\tilde{\Psi}| 
\;. 
\label{eq1} 
\end{equation} 
Furthermore, the ket and bra states are parametrised within the single-reference CCM as follows:   
\begin{eqnarray} 
|\Psi\rangle = {\rm e}^S |\Phi\rangle \; &;&  
\;\;\; S=\sum_{I \neq 0} {\cal S}_I C_I^{+}  \nonumber \; , \\ 
\langle\tilde{\Psi}| = \langle\Phi| \tilde{S} {\rm e}^{-S} \; &;& 
\;\;\; \tilde{S} =1 + \sum_{I \neq 0} \tilde{{\cal S}}_I C_I^{-} \; .  
\label{eq2} 
\end{eqnarray} 
One of the most important features of the CCM is that one uses a 
single model or reference state $|\Phi\rangle$ that is normalised.
We note that the parametrisation of the ground state has the normalisation condition for the 
ground-state bra and ket wave functions ($\langle \tilde\Psi|\Psi\rangle 
\equiv \langle\Phi|\Phi\rangle=1$). The model state is required to have the 
property of being a cyclic vector with respect to two well-defined Abelian 
subalgebras of {\it multi-configurational} creation operators $\{C_I^{+}\}$ 
and their Hermitian-adjoint destruction counterparts $\{ C_I^{-} \equiv 
(C_I^{+})^\dagger \}$.  The interested reader is referred to the Appendices
and to Ref. \cite{ccm39} for more information regarding how the CCM problem 
is solved for.

Here, we use the N\'eel state as the model state for the antiferromagnetic Heisenberg model
given by
\begin{equation}
H = J \sum_{\langle i,j \rangle}^N \; \; {\bf s}_i \cdot {\bf s}_{j }  \; \; .
\label{H}
\end{equation}
For the bipartite lattices, we perform a rotation of the local axes
of the up-pointing spins by 180$^\circ $ about the
$y$-axis. The transformation is described by,
\begin{equation}
s^x \; \rightarrow \; -s^x, \; s^y \; \rightarrow \;  s^y, \;
s^z \; \rightarrow \; -s^z  \; .
\end{equation}
The model state now appears $mathematically$ to consist
of purely down-pointing spins. 
In terms of the spin raising and lowering operators
$s_k^{\pm} \equiv s_k^x \pm {\rm i} s_k^y$ the Hamiltonian 
may be written in these local axes as,
\begin{equation}
H = - J \sum_{\langle i,j \rangle}^N \; \biggl[ \; s_i^+
s_j^+ + s_i^-s_{j }^- + 2 s_i^z s_{j }^z  \; \biggr] \; ,
\label{eq:newH}
\end{equation}
where the sum on $\langle i,j\rangle$ again counts all nearest-neighbour pairs once
on the lattice. 

The CCM formalism is exact in the limit of inclusion of
all possible multi-spin cluster correlations within 
$S$ and $\tilde S$, although this is usually impossible to achieve
practically.  
Hence, we generally make approximations in both $S$ and $\tilde S$.  The three most commonly employed approximation schemes previously utilised have been: (1) the SUB$n$ scheme, in which all correlations involving only $n$ or fewer spins are retained, but no further restriction is made concerning their spatial separation on the lattice; (2) the SUB$n$-$m$  sub-approximation, in which all SUB$n$ correlations spanning a range of no more than $m$ adjacent lattice sites are retained; and (3) the localised LSUB$m$ scheme, in which all multi-spin correlations over all distinct locales on the lattice defined by $m$ or fewer contiguous sites are retained. 
Another important feature 
of the method is that the bra and ket states are not always 
explicitly constrained to be Hermitian conjugates when we 
make such approximations, although the important 
Helmann-Feynman theorem is always preserved. We remark 
that the CCM provides results in the infinite-lattice limit 
$N \rightarrow \infty$ from the outset.

Key to understanding the application of the CCM to spin problems is the
concept of the unit cell and the Bravais lattice. The unit cell
contains a number of sites at specific positions (given by the ``primitive'' lattice
vectors) that are replicated at all possible multiples of the Bravais lattice
vectors. Thus, for example, we have a single site in the unit cell for the linear
chain, say, at position $(0,0)$ and a single Bravais lattice vector $\hat a=(1,0)^T$.
The lattice is formed by translating the single site in the unit cell by all integer
multiples of $\hat a$. Two-dimensional lattices have two Bravais lattice vectors. For example
the square lattice has a single site in the unit cell and the lattice vectors
are $\hat a = (1,0)^T$ and  $\hat b=(0,1)^T$. The triangular lattice is given by
vectors $\hat a=(1,0)^T$ and $\hat b=(1/2,\sqrt{3}/2)^T$ and so on for other lattices.
We see also that the basic building blocks of unit cell, Bravais lattice, and
bonds/interactions in the Hamiltonian placed on the lattice gives us a broad
canvas to work with. For example, we may form models that interpolate between different
lattices (and even different spatial dimensions) by varying the strengths of various bonds
that have been carefully placed with respect to the underlying lattice. Hence, the number of
possible such quantum spin systems is enormous. Furthermore, the development in the
number and complexity of these theoretical models is often driven by the magnetic materials
studied in experiment.

\section{Results}

\subsection{The Spin-Half $J_1$--$J_2$ Model on the Linear Chain}

The Hamiltonian for this spin-half model has nearest-neighbour bonds of 
strength $J_1$ and next-nearest-neighbour bonds of strength $J_2$. We use 
a N\'eel model state in which nearest-neighbour spins on the linear chain are anti-parallel. We rotate the spin coordinates of the `up' spins so that notationally they become `down' spins in these locally defined axes. The relevant Hamiltonian in rotated coordinates is then given by
\begin{equation}
H = -J_1 \sum_{\langle i,j\rangle} \bigl ( s_i^z s_j^z + \frac 12 \{  s_i^- s_j^- + s_i^+ s_j^+ \} \bigr )  
       + 
        J_2 \sum_{\langle \langle i,k \rangle \rangle} \bigl ( s_i^z s_k^z + \frac 12 \{  s_i^+ s_k^- + s_i^- s_k^+ \} \bigr ) ~~ ,
\label{j1j2hamiltonian}
\end{equation}
where $\langle i,j\rangle$ runs over all nearest-neighbour sites on the lattice counting each pair once and once only and $\langle \langle i,k \rangle \rangle$ runs 
over all next-nearest-neighbor sites on the lattice, again counting each pair once and once only. Henceforth we put $J_1=1$ and consider $J_2>0$. 

The ground-state properties of this system have been studied using methods such as 
exact diagonalisations  \cite{mg2,aligia00}, DMRG  \cite{mg3,mg4,mg5,ccm7}, CCM
 \cite{ccm4,ccm6,ccm29}, and field-theoretical approaches   \cite{mg5} (see Refs. 
 \cite{mg5,mg6} for a general review).  We shall not go into detail about this
 model here except to note that there are two degenerate  simple exact dimer-singlet 
 product ground states at the Majumdar-Ghosh 
point $J_2/J_1=0.5$.

We now consider how this model can be treated at the Majumdar-Ghosh 
point $J_2/J_1=0.5$by the CCM via  the identification of a  special dimerised 
solution of the CCM equations for a N\'eel model state. 
We use a doubled unit cell including two neighbouring sites for a spin-half 
system on the linear chain at points (0,0,0) and (1,0,0) and 
a single Bravais vector (2,0,0)$^T$ to take into account the symmetry breaking. 
There are thus two distinct types of two-spin 
nearest-neighbour ket-state correlation coefficients and again 
these are denoted as ${\cal S}_2^a$  and ${\cal S}_2^b$ 
for $S_2 = {\cal S}_2^a \sum_{i_a} s_{i_a}^+ s_{i_a+1}^+ 
+ {\cal S}_2^b \sum_{i_b} s_{i_b}^+ s_{i_b+1}^+$ with respect to the CCM 
ket state of Eq. (\ref{eq2}), and where $i_a$ runs over all sites with 
odd-numbered indices and $i_b$ runs over all sites with even-numbered 
indices. The exact ground state at $J_2/J_1=0.5$  is 
obtained by setting ${\cal S}_2^a=1$ and all other coefficients equal to zero. 
This result is found to hold for all levels of LSUB$m$ approximation. 
The exact ground-state energy of $E_g/N=-0.375 J_1$ is thus
obtained at the point $J_2/J_1=0.5$, as expected. Furthermore, 
the sublattice magnetisation is found to be zero at this point at all
levels of approximation. 
We find values \cite{ccm41} for the excitation energy gap of $0.35250$, $0.34170$, 
$0.30548$, $0.28732$, $0.27559$, and $0.26760$ at the LSUB$m$ 
levels of approximation with $m=\{4,6,8,10,12,14\}$.  A simple extrapolation
of these results in the limit $m \rightarrow \infty$ using a quadratic function
gives a value for the gap of 0.2310. This result is in agreement with 
results of exact diagonalisations that predict a gap of 0.234 \cite{caspers}. 

\subsection{The $J$--$J'$ Heisenberg antiferromagnet on the CAVO lattice}
\label{cavo_n_n}

\begin{figure}
\epsfxsize=7cm
\centerline{\epsffile{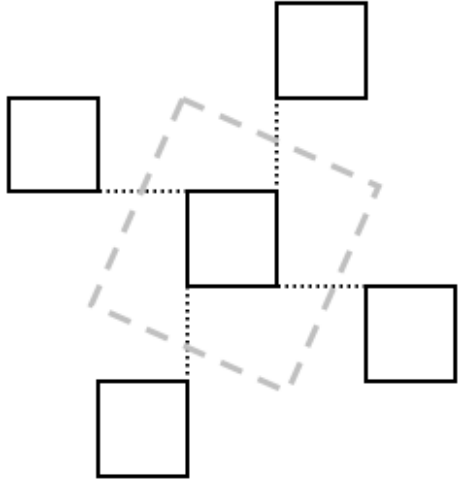}}
\caption{The CAVO lattice.
The nearest-neighbour bonds that connect two sites 
on a four-site plaquette are shown by the solid 
lines and have a bond strength 
given by $J$ (=1). The nearest-neighbour bonds that connect two sites 
on different plaquettes (dimer bonds) are shown by the dotted lines 
and have a bond strength 
given by $J'$. The unit cell of the lattice is shown by the square with the grey dashed lines.}
\label{cavo_unit_cell}
\end{figure}
 
In the previous section, we saw that the exact dimerised state at the 
Majumdar-Ghosh point may be obtained for the N\'eel model state 
by appropriate choice of ket-state correlation coefficients. However,
this is a case in which the translational symmetry of the lattice (the
linear chain) is spontaneously broken. We now wish to a
case in which dimerised and plaquette exact solutions follow the 
symmetry of the Hamiltonian and so do not form symmetry-broken states.
We consider an antiferromagnetic Heisenberg model 
in which the basic geometric unit cell 
contains four neighbouring lattice sites on the underlying 
crystallographic lattice of the 
magnetic material CaV$_4$O$_9$ (CAVO), shown in Fig. \ref{cavo_unit_cell}.
There are two non-equivalent antiferromagnetic 
nearest-neighbour bonds $J$ and $J'$ 
belonging to dimers ($J'$) and to 
four-spin plaquettes ($J$) respectively.  The ground state of the quantum 
model depends on the ratio $J'/J$ of the competing bonds.  
Using a unit cell as defined in Fig.~\ref{cavo_unit_cell}, the plaquette bonds
$J$ are inside the four-site unit cell and the dimer bonds $J'$ connect sites in different unit
cells. We note that this model is not frustrated but the two  non-equivalent
nearest-neighbour bonds lead to a competition in the quantum system.  
Henceforth, we choose an energy scale such that $J=1$.

The four-site plaquettes in the unit cell become de-coupled only in the limit
$J'/J \rightarrow 0$.  The ground state is a product of such four-site plaquette singlets 
in this limit. In the limit that $J'/J \rightarrow \infty$ dimers are formed on the 
$J'$ bonds. To model such states using the CCM we start again from the N\'eel model state; 
namely, a state in which the spins on nearest-neighbour sites are anti-parallel. 
To create an exact plaquette-singlet product VBC ground state at $J'/J=0$ 
using the CCM we have to adjust the 
nearest-neighbour correlation coefficients ${\cal S}_2^a$ and  ${\cal S}_2^b$
and a single four-body plaquette correlation  
coefficient ${\cal S}_4^p$ containing all four sites properly. 
(Note that  ${\cal S}_2^a$ represents those ket-state coefficients for 
the nearest-neighbour two-body cluster connecting sites on a plaquette 
indicated by the solid lines in Fig. \ref{cavo_unit_cell}, 
whereas ${\cal S}_2^b$ represents those ket-state coefficients 
for the nearest-neighbour two-body cluster connecting sites 
on a dimer indicated by the dotted lines in the same figure. 
The coefficient ${\cal S}_4^p$ represents those ket-state coefficients for 
the four-body cluster corresponding to a plaquette 
indicated by the solid lines in Fig. \ref{cavo_unit_cell}.)
Indeed, it is easy to show that setting the ket-state correlation 
coefficients ${\cal S}_2^a$ and ${\cal S}_4^p$ to a value of 0.5 and all 
other ket-state correlation coefficients (including ${\cal S}_2^b$) to zero the  
plaquette-singlet product valence-bond crystal state is obtained exactly. 
Furthermore, we are also able 
to reproduce exactly the  dimer-singlet product ground state 
in the limit $J'/J \rightarrow \infty$. In this limit, the nearest-neighbour 
ket-state correlation coefficient ${\cal S}_2^b$ on 
the dimer bonds (dotted lines in Fig.
\ref{cavo_unit_cell}) has a value of one 
and all other coefficients (e.g., ${\cal S}_2^a$ and
${\cal S}_4^p$) are zero. 

\begin{figure}
\epsfxsize=11cm
\centerline{\epsffile{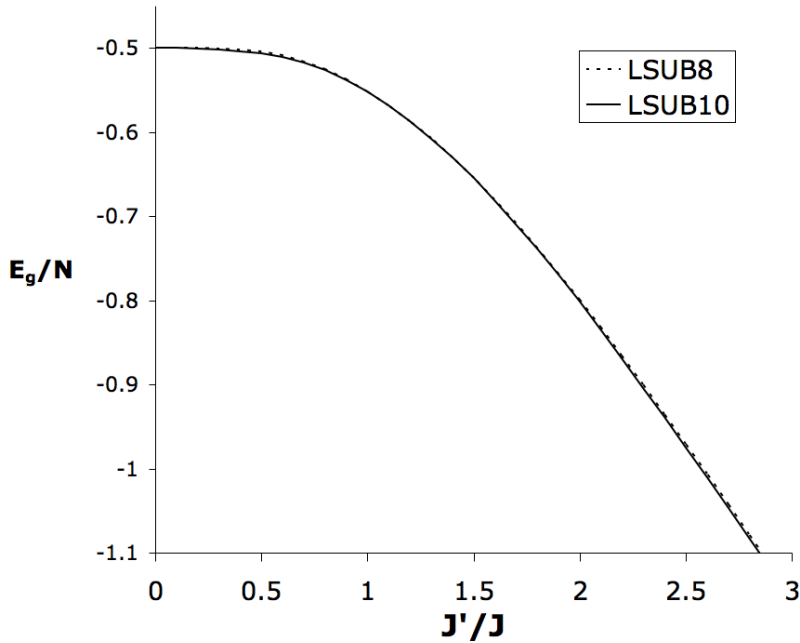}}
\caption{CCM results for the ground-state energy of the $J$--$J'$ 
Heisenberg antiferromagnet on the CAVO lattice (with $J=1$).}
\label{cavo_energies}
\end{figure}

An important point is that in the limits $J'/J \rightarrow 0$ and $|J'/J| \rightarrow \infty$
the system is comprised of independent clusters. However, the system is two-dimensional
for all other values of $J'/J$. This system therefore ``interpolates'' between a zero-dimensional 
and two-dimensional lattice with the bond strengths $J'/J$. This model (and similar models) may 
therefore be used to study the differences between zero-dimensional  and  two-dimensional systems. 
They may also be used to investigate the effects of ``linking'' magnetic clusters in order to 
form an extended two-dimensional material. Such models are of interest in the subject of 
quantum computational.  We remark that the CCM solution at intermediate values of  $J'/J$ 
may be found by ``tracking'' the exact CCM solution by making small incremental
changes in $J'/J$ from either of the exact limits $J'/J \rightarrow 0$ and $|J'/J| \rightarrow \infty$ 
(see Ref. \cite{ccm37}  for details).

We remark that the correct ground-state energies (with $J=1$) are reproduced 
in the limits $J'/J \rightarrow 0$ and $J'/J \rightarrow \infty$, namely, 
$E_g/N=-0.5$ and $E_g/N=-0.375J'$, respectively. These results for the ground-state 
energy are shown in Fig. \ref{cavo_energies}, and we see that LSUB8 and LSUB10
results agree extremely well over the region considered. Hence, we conclude that the 
CCM also provides excellent results for intermediate values of $J'/J$ that interpolates 
between the two (exact) limits, $J'/J \rightarrow 0$ and $J'/J \rightarrow \infty$.

\subsection{The Spin-Half and Spin-One Heisenberg Model on Finite-Sized Chains}

We have seen in the previous sections that an appropriate choice of the unit cell 
and values for the CCM  correlation coefficients can reproduce exact results for
N\'eel model states. This was observed, namely, for the symmetry-breaking solution 
to the $J_1$--$J_2$ model at $J_2/J_1=0.5$ or for the CAVO model with two
types of bonds in the limits $J'/J \rightarrow 0$ and $|J'/J| \rightarrow \infty$. 
We now consider if exact solutions for larger unit cells may be found by ``brute force'',
i.e., without using any of the translational or point-group symmetries of the  finite
lattice.

We start by defining the unit cell to be of size $N=\{2,4,6,8,10,12\}$. We use the 
Hamiltonian of Eq. (\ref{H}) and we rotate the ``up'' spins as usual. The Hamiltonian 
in the new spin coordinates is then given by Eq. (\ref{eq:newH}). Hence, we form 
bonds (of strength $J=1$) between all nearest-neighbour sites in the unit cell. 
However, we do not ``link'' the bonds by forming an intermediate bond between 
the unit cells, and so, in effect, the unit cells become isolated clusters. A technical 
point is that we set the single Bravais lattice vector to be of magnitude $+N$ for
this ``brute force'' approach;
again, even though the clusters/unit cells are not linked by intermediate bonds. 
Furthermore, we may also provide an additional bond between the first and last sites (again 
of strength $J=1$). In effect, this bond acts as a ``boundary condition'' that creates
a ``ring'' of sites for the finite lattice. These results should be compared to those
of ED with periodic boundary conditions.

High-order CCM SUB$n$-$m$ results for the antiferromagnetic Heisenberg model 
(with periodic conditions) for finite-sized chains of length $N$ with SUB$n$-$m$ with $n=m=N$ 
for $s=1/2$ and $n/2=m=N$ for $s=1$ in Tables \ref{tab1} and \ref{tab2}, respectively. 
We remark that CCM results for the ground-state energy and excitation energy
gap agree to at least six decimal places with exact diagonalisations obtained
using the SpinPack code of Joerg Schulenburg \cite{spinpack}.
We may also study the manner in which LSUB$m$ results behave with increasing $m$
for a set value of chain length $N$. Table \ref{tab3} presents LSUB$m$ results for the 
spin-half  Heisenberg chain of  length $N=12$. The ground-state energies decrease 
monotonically with increasing LSUB$m$ level of approximation, although no simple
extrapolation ``rule'' may be seen. By contrast, the LSUB$m$ data for the excitation 
energy gap is only monotonically decreasing up to LSUB10. Indeed, the gap for
LSUB10 lies lower than that of the (exact) LSUB12 result. We note again that the LSUB$m$
approximation for the spin-half system and the SUB$2m$-$m$ approximation
for the spin-one system reproduce exact results for the chains of length $N$ when
we set $m=N$. 

The CCM results in Tables  \ref{tab1} to \ref{tab3} arise from ``special solutions'' of the CCM 
equations in exactly the same manner as for the nearest-neighbour CAVO problem 
in the limits $J'/J \rightarrow 0$ and $|J'/J| \rightarrow \infty$. However, 
it is clearly more complicated in this case because we are dealing with 
unit cells (i.e., finite-sized 1D lattices here) of greater size than 4 sites. 
Interestingly though, the CCM ground-state equations were found to converge 
readily to this ``special solution'' and for a wide range of starting values at all 
levels of SUB$n$-$m$ approximation attempted here. For example, setting the
initial ket-state correlation coefficients to a small (positive) non-zero value
was found to work quite well. Indeed, the ground-state solution 
was found to be very stable, as is generally also the case for extended systems 
($N \rightarrow \infty$) in which the model state is known to be a ``good
starting point''. 

These results {\it prove the principle} that the CCM may be 
used to study finite-sized lattices.

\begin{table}[t]
\caption{CCM LSUB$m$ results for the ground-state energy and excitation energy gap 
of spin-half antiferromagnetic Heisenberg chains of $N$ length with periodic boundary 
conditions.}
\begin{center}
{\footnotesize
\begin{tabular}{|c|c|c|c|c|}                                                                                     
\hline\noalign{\smallskip}
N=  &CCM $E_g/N$    &CCM Gap                 \\ \hline \hline 
4     &$-$0.5                   &1                     \\
6     &$-$0.46712928   &0.684740       \\
8     &$-$0.45638668   &0.522676       \\
10   &$-$0.45154464   &0.423239        \\
12   &$-$0.44894924   &0.355848      \\
\noalign{\smallskip}\hline
\end{tabular}
}
\end{center}
\label{tab1}
\end{table}

\begin{table}[t]
\caption{CCM SUB$n$-$m$ results (with $n=2N$ and $m=N$) 
for the ground-state energy and excitation energy gap 
of spin-one antiferromagnetic Heisenberg chains of $N$ length with periodic boundary 
conditions.}
\begin{center}
{\footnotesize
\begin{tabular}{|c|c|c|c|c|}                                                                                     
\hline\noalign{\smallskip}
N=  &$E_g/N$                 &Gap           \\ \hline \hline 
2     &$-$1.263853992   &1.4876903286     \\
4     &$-$1.29781459     &1.038539433      \\
6     &$-$1.436237197   &0.720627363       \\
\noalign{\smallskip}\hline
\end{tabular}
}
\end{center}
\label{tab2}
\end{table}

\begin{table}[t]
\caption{CCM LSUB$m$ results for the ground-state energy and excitation energy gap 
of spin-half antiferromagnetic Heisenberg chains of length $N=12$ with periodic boundary 
conditions.}
\begin{center}
{\footnotesize
\begin{tabular}{|c|c|c|c|c|}                                                                                     
\hline\noalign{\smallskip}
LSUB$m$ $m=$ &$E_g/N$                   &Gap   \\ \hline \hline 
4                             &$-$0.421064786    &0.589781571   \\
6                             &$-$0.42329432      &0.440170513   \\
8                             &$-$0.423975557    &0.375801129   \\
10                           &$-$0.424415368    &0.348833726   \\
12                           &$-$0.448949243    &0.355847514   \\
\noalign{\smallskip}\hline
\end{tabular}
}
\end{center}
\label{tab3}
\end{table}

\section{Conclusions}

Again, we remark that recent developments of high-order CCM have been to extend 
existing formalism and codes to $s \ge \frac 12$ for both the ground and excited 
states \cite{ccm39}, and independently to ``generalised'' expectation values for a 
wide range of one- and two-body spin operators \cite{ccm40}. We note that the 
CCM is that the Goldstone linked-cluster theorem is obeyed at all levels of 
approximation and so it provides results in the infinite lattice limit $N \rightarrow 
\infty$ from the outset. 
In this article, we have shown the exact ground state for the $J_1$--$J_2$ model 
at $J_2/J_1=0.5$ may be reproduced using a N\'eel model state. Using a similar
model state in 2D, exact results for a nearest-neighbour CAVO model were reproduced
using high-order CCM in the limits $J'/J \rightarrow 0$ (4-site plaquette) and 
$J'/J \rightarrow \infty$ (2-site dimers). These results lead on naturally to  a 
``brute-force'' approach for solving finite-sized lattices, i.e., without using any of the 
translational or point-group symmetries of the finite lattice. Indeed, we have
shown here that high-order CCM SUB$n$-$m$ results for the antiferromagnetic 
Heisenberg model (with periodic conditions) for finite-sized chains of length $N$ 
with SUB$n$-$m$ with $n=m=N$ for $s=1/2$ and $n/2=m=N$ for $s=1$ were 
found to agree with exact-diagonalisation (ED) results for the ground-state energy 
and excitation energy gap agree to at least six decimal places.

We note that results for the case of finite clusters for $s=1$ were obtained via 
new high-order formalism for the CCM excited state outlined in Refs. \cite{ccm39,ccm41}. 
Furthermore, new code \cite{thecode} has been written to implement this
new formalism for the excited state for $s \ge 1$. The agreement between 
ED results for the ground-state energy and excitation energy gap for chains of 
up to $N=6$ for $s=1$ is an excellent test of the validity of this new code. These
solutions for the finite chains were found to be stable numerically. 
Hence, we have ``proven the principle'' that the high-order CCM code may be used 
directly to study (relatively small-sized) finite lattices by a somewhat ``brute-force'' 
approach. Indeed, this approach is still somewhat inefficient in 
comparison to exact diagonalisations (ED) because ED results use translational 
and point-group symmetries of the finite lattice in order to reduce the size of the 
matrix to be diagonalised, and so ED may to go to much larger lattice sizes.

For periodic boundary conditions, the manner in which this is achieved for ED 
is by identifying common states for a given $k$-value via a complex phase factor 
$e^{-{\rm i} {\bf k} \cdot {\bf r}}$ for states that a related by a translational vector 
$r$ on the finite lattice (again: note periodic boundary conditions are assumed). 
The Hamiltonian then links only those states of common $k$. Point-group (PG) 
symmetries/permutations of indices may then used to form states with {\it real} 
components only, thus simplifying the computational problem. 
In principle, we ought to be able to employ the finite lattice translational and point-group 
symmetries analogously in order to simplify the finite-size problem with periodic 
boundary conditions also for the CCM. However, it is unclear how one might do this
in practice for the exponentiated $S$ in the ground ket and bra states. 
The use of translational and PG symmetries for finite-lattice CCM will be the 
subject of future research. 

Despite the fact that translational and PG symmetries for the finite lattice were not
used directly here, we were still able to treat finite-lattices of size $N=12$ for
the spin-half case using high-order CCM with only relatively meagre computational
resources. (A MacBook with a ``Core Duo'' processor and 1 GIG RAM was used in this 
case was used to carry out this calculation.) Still larger lattices are possible using 
the CCCM code \cite{thecode}, which has been implemented to work in parallel on a 
cluster of processors. We note that the addition of intermediate bonds that link the 
isolated clusters in order to form extended lattices of infinite numbers of sites requires 
only relatively small increases in computational effort for high-order CCM 
compared to treating the case of isolated magnetic clusters alone. Excellent
results were seen here for such a model that interpolated between finite
clusters and an extended lattice, namely, for the n.n. CAVO model 
(e.g., see the results for the ground state energy in Fig. \ref{cavo_energies}). 
High-order CCM might provide a good choice for the study of a whole range of 
such ``interpolating'' models between finite clusters and infinite lattices.



\begin{thebibliography}{200}


\bibitem{refc1} F. Coester, Nucl. Phys.  {\bf 7}, 421 (1958);  F. Coester and H. K\"ummel, {\em ibid.} {\bf  17}, 477 (1960).
\bibitem{refc2} J. \v{C}i\v{z}ek, J. Chem. Phys.  {\bf 45}, 4256 (1966);  Adv. Chem. Phys.  {\bf 14}, 35 (1969).
\bibitem{refc3} R.F. Bishop and K.H. L\"uhrmann, Phys. Rev. B  {\bf 17}, 3757 (1978); {\it ibid.}  {\bf 26}, 5523 (1982). 
\bibitem{refc4} H. K\"ummel, K.H. L\"uhrmann, and J.G. Zabolitzky, Phys Rep.  {\bf 36C}, 1 (1978).
\bibitem{refc5} J.S. Arponen, Ann. Phys. (N.Y.)  {\bf 151}, 311 (1983).
\bibitem{refc6} R.F. Bishop and H. K\"ummel, Phys. Today  {\bf 40(3)}, 52 (1987).
\bibitem{refc7} J.S. Arponen, R.F. Bishop, and E. Pajanne, Phys. Rev. A  {\bf 36}, 2519 (1987);		
{\em ibid.}  {\bf 36}, 2539 (1987); in: {\em Condensed Matter Theories}, Vol.  {\bf 2},                 
P. Vashishta, R.K. Kalia, and R.F. Bishop, eds. (Plenum, New York, 1987), 		p. 357.
\bibitem{refc8} R.J. Bartlett, 	J. Phys. Chem.  {\bf 93}, 1697 (1989).
\bibitem{refc9} R.F. Bishop, Theor. Chim. Acta  {\bf 80}, 95 (1991). 
\bibitem{ccm1} M. Roger and J.H. Hetherington,    Phys. Rev. B  {\bf 41}, 200 (1990);   Europhys. Lett.  {\bf 11}, 255 (1990).
\bibitem{ccm2} R.F. Bishop, J.B. Parkinson, and Y. Xian,   Phys. Rev. B  {\bf 44}, 9425 (1991).
\bibitem{ccm999} R.F. Bishop, J.B. Parkinson, and Y. Xian,  Phys. Rev. B  {\bf 46}, 880 (1992). 
\bibitem{ccm3} R.F. Bishop, J.B. Parkinson, and Y. Xian,    J. Phys.: Condens. Matter {\bf  5}, 9169 (1993).
\bibitem{ccm4} D.J.J. Farnell and J.B. Parkinson,   J. Phys.: Condens.  Matter  {\bf 6}, 5521 (1994). 
\bibitem{ccm5} R.F. Bishop, R.G. Hale, and Y. Xian,  Phys. Rev. Lett.  {\bf 73}, 3157 (1994).
\bibitem{ccm6} Y. Xian, J. Phys.: Condens.\ Matter    {\bf 6}, 5965 (1994). 
\bibitem{ccm7} R. Bursill, G.A. Gehring, D.J.J. Farnell, J.B. Parkinson, T.   Xiang, and C. 
Zeng, J. Phys.: Condens. Matter  {\bf 7}, 8605 (1995). 
\bibitem{ccm8} R.G. Hale. Ph.D. Thesis, UMIST, Manchester, United Kingdom (1995).
\bibitem{ccm9} R.F. Bishop, D.J.J. Farnell, and J.B. Parkinson,   J. Phys.: Condens. Matter  {\bf 8}, 11153 (1996).
\bibitem{ccm10} D.J.J. Farnell, S.A. Kr\"uger, and J.B. Parkinson,   J. Phys.: Condens. Matter  {\bf 9}, 7601 (1997).
\bibitem{ccm11} R.F. Bishop, Y. Xian, and C. Zeng, in:  {\it Condensed   Matter Theories}, Vol.  {\bf 11}, 
E.V. Lude\~na, P. Vashishta,  and R.F. Bishop, eds. (Nova Science, Commack, New York, 1996),   p. 91.
\bibitem{ccm12} C. Zeng, D.J.J. Farnell, and R.F. Bishop,   J. Stat. Phys.  {\bf 90}, 327 (1998).
\bibitem{ccm13} R.F. Bishop, D.J.J. Farnell, and J.B. Parkinson,   Phys. Rev. B   {\bf 58}, 6394 (1998).
\bibitem{ccm13a} R.F. Bishop
in 
{\it Microscopic Many-Body Theories and
                   Their Applications}, Lecture Notes in Physics   {\bf 510},
J. Navarro and A. Polls, eds.
                   Lecture Notes in Physics Vol.  {\bf 510} (Springer-Verlag,
                   Berlin, 1998), p. 1.
\bibitem{ccm14} J. Rosenfeld, N.E. Ligterink, and R.F. Bishop,  Phys. Rev. B {\bf  60}, 4030 (1999).
\bibitem{ccm15} R.F. Bishop, D.J.J. Farnell, S.E. Kr\"uger, J.B.  Parkinson, 
J. Richter, and C. Zeng, J. Phys.: Condens. Matter  {\bf 12}, 6887 (2000).
\bibitem{ccm16}  R.F. Bishop, D.J.J. Farnell, and M.L. Ristig,  Int. J. Mod. Phys. B  {\bf 14}, 1517 (2000).
\bibitem{ccm17} S.E. Kr\"uger, J. Richter, J. Schulenberg,  D.J.J. Farnell, and R.F.
Bishop, Phys. Rev. B  {\bf 61}, 14607 (2000). 
\bibitem{ccm18} D.J.J. Farnell, R.F. Bishop, and  K.A. Gernoth,   Phys. Rev. B  {\bf 63}, 220402R (2001). 
\bibitem{ccm19} D.J.J. Farnell, K.A. Gernoth, and R.F. Bishop,  Phys. Rev. B    {\bf 64}, 172409 (2001). 
\bibitem{ccm19a} S.E. Kr\"uger and J.Richter,
       Phys. Rev. B  {\bf 64}, 024433 (2001).

\bibitem{ccm20} D.J.J. Farnell, R.F. Bishop, and K.A. Gernoth,  J. Stat. Phys.  {\bf 108}, 401 (2002).
\bibitem{ccm21} N.B. Ivanov, J. Richter, and D.J.J. Farnell, Phys.  Rev. B   {\bf 66}, 014421 (2002).
\bibitem{ccm22} D.J.J. Farnell and R.F. Bishop, 
{ arxiv.org/abs/cond-mat/0311126}.
\bibitem{ccm23} S.E. Kr\"uger, D.J.J. Farnell, and J. Richter, 
Int. J. Mod. Phys. B {\bf 17}, 5347 (2003).
 
 \bibitem{ccm24}  R. Darradi, J. Richter, and D.J.J. Farnell, Phys. Rev. B. {\bf 72}, 104425 (2005).

\bibitem{ccm24a} R. Darradi, J. Richter, and D.J.J. Farnell, J. Phys.: Condens. Matter {\bf 17}, 
341 (2005).
 
\bibitem{ccm26}  D.J.J. Farnell, J. Schulenberg, J. Richter, and K.A. Gernoth, Phys. Rev. B. {\bf 72} , 172408 (2005).
\bibitem{ccm27} S.E. Kr\"uger, R. Darradi, J. Richter, and D.J.J Farnell, Phys. Rev. B {\bf 73}, 094404 (2006)
\bibitem{ccm27a}  D. Schmalfu{\ss}, R. Darradi, J. Richter, J.~Schulenburg,  and
D.~Ihle, Phys. Rev. Lett.  {\bf 97}, 157201 (2006).

\bibitem{ccm28} D.J.J. Farnell and R.F. Bishop, { arxiv.org/abs/cond-mat/0606060}.
\bibitem{ccm29} J. Richter, R. Darradi,  R. Zinke, and R.F. Bishop,
       Int. J. Mod. Phys. B  {\bf 21}, 2273 (2007).
\bibitem{ccm30} R. Zinke,  J. Schulenburg, and J. Richter,
      Eur. Phys. J. B  {\bf 61}, 147 (2008).
\bibitem{ccm31} R.F. Bishop, P.H.Y. Li, R. Darradi, and J.Richter,
J. Phys.: Condens. Matt.  {\bf 20} 255251 (2008).
\bibitem{ccm32} R.F. Bishop, P.H.Y. Li, R. Darradi, J.~Schulenburg, and J.Richter,
Phys. Rev. B {\bf 78}, 054412 (2008).
\bibitem{ccm33} R.F. Bishop, P.H.Y. Li, R. Darradi, and J.Richter,
Europhys. Lett. {\bf 83}, 47004 (2008).  
\bibitem{ccm34} R.F. Bishop, P.H.Y. Li, R. Darradi, J.Richter, and C.E. Campbell, 
J. Phys.: Condens. Matt. {\bf 20}, 415213 (2008).
\bibitem{ccm35} 
R. Darradi, O. Derzhko, R. Zinke, J. Schulenburg,
S.E. Kr\"uger, and J. Richter, Phys. Rev. B {\bf 78}, 214415 (2008).
  

\bibitem{ccm36} D.J.J. Farnell and R.F. Bishop,  Int. J. Mod. Phys. B. {\bf 22}, 3369 (2008).

\bibitem{ccm37} D.J.J. Farnell, J. Richter, R. Zinke, and R.F. Bishop, J. Stat. Phys. {\bf 135}, 175 (2009). 

\bibitem{ccm38} P. Li, D.J.J. Farnell, and R.F. Bishop, Phys. Rev. B {\bf 79}, 174405 (2009). 

\bibitem{ccm39} D.J.J. Farnell, arXiv:0909.1226

\bibitem{ccm40} D.J.J. Farnell, arXiv:0911.5150

\bibitem{ccm41} D.J.J. Farnell, in {\it Condensed Matter Theories} -- in print.

\bibitem{thecode} A GPL licensed version of the `Crystallographic Coupled Cluster Method' (CCCM) code of D.J.J. Farnell and J.~Schulenburg is available online at: http://www.ovgu.de/jschulen/ccm/ 



\bibitem{mg1} C.K Majumdar and D.K. Ghosh, J. Math. Phys.  {\bf 10}, 1388 (1969);  J. Math. Phys.  {\bf 10}, 1399 (1969). 
\bibitem{mg2} T. Tonegawa and I. Harada, J. Phys. Soc. Japan {\bf  56}, 2153 (1987). 
\bibitem{mg3}  K. Nomura and K. Okamoto,  Phys. Lett. {\bf  169A}, 433 (1992); J. Phys. Soc. Japan {\bf  62}, 1123 (1993);  
J.Phys. A.: Math. Gen.   {\bf 27}, 5773 (1994). 
%
\bibitem{mg4} R. Chitra, S. Pati, H.R. Krishnamurthy, D. Sen, and S. Ramasesha, Phys. Rev. B  {\bf 52}, 6581 (1995). 
\bibitem{mg5} S. R. White and I. Affleck, Phys. Rev. B {\bf 54}, 9862 (1996).
\bibitem{mg6}  H.-J. Mikeska and A. K. Kolezhuk, in {\it Quantum Magnetism}, 
Lecture Notes in Physics  {\bf 645}, U. Schollw\"{o}ck, J. Richter, D.J.J. Farnell, and R.F. Bishop, eds.
(Springer-Verlag, Berlin, 2004), pp 1-83. 
\bibitem{aligia00} 
A.A.~Aligia, C.D.~Batista, and F.H.L.~E{\ss}ler, 
{ Phys. Rev. B}  {\bf 62}, 3259 (2000). 

\bibitem{caspers} W.J. Caspers, in: {\it Spins Systems} (World Scientic, 1989) p. 107.
29


\bibitem{spinpack} The exact diagonalisations {\it SpinPack} code of J. Schulenburg is available under GPL licence at: http://www.ovgu.de/jschulen/spin/


\end{thebibliography}
\end{document}